\documentclass[11pt,a4paper]{article}
\usepackage{jcappub}
\usepackage{bm}
\usepackage{indentfirst}
\usepackage{amsmath}
\usepackage{graphicx}
\usepackage{float}
\usepackage{amssymb}
\usepackage{subfigure}

\begin{document}
\title{Primordial Non-Gaussianity from G-inflation}
\author[a]{Fengge Zhang,}
\author[a,b,1]{Yungui Gong\note{Corresponding author.},}
\author[a]{Jiong Lin,}
\author[a]{Yizhou Lu,}
\author[c]{Zhu Yi}

\affiliation[a]{School of Physics, Huazhong University of Science and Technology,
Wuhan, Hubei 430074, China}
\affiliation[b]{MOE Key Laboratory of TianQin Project, Sun Yat-sen University, Zhuhai 519082, China}
\affiliation[c]{Department of Astronomy, Beijing Normal University,
Beijing 100875, China}

\emailAdd{fenggezhang@hust.edu.cn}
\emailAdd{Corresponding author. yggong@hust.edu.cn}
\emailAdd{jionglin@hust.edu.cn}
\emailAdd{louischou@hust.edu.cn}
\emailAdd{yz@bnu.edu.cn}

\abstract{
Enormous information about interactions is contained in the non-Gaussianities of the primordial curvature perturbations, which are essential to break the degeneracy of inflationary models.
We study the primordial bispectra for G-inflation models predicting  both sharp and broad peaks in the primordial scalar power spectrum. 
We calculate the non-Gaussianity parameter $f_{\mathrm{NL}}$ in the equilateral limit and squeezed limit numerically, 
and confirm that the consistency relation holds in these models.
Even though $f_{\mathrm{NL}}$ becomes large at the scales before the power spectrum reaches the peak and the scales where there are wiggles in the power spectrum, 
it remains to be small at the peak scales.
Therefore, the contributions of non-Gaussianity to the scalar induced secondary gravitational waves and primordial black hole abundance are expected to be negligible.}

\arxivnumber{2012.06960}

\maketitle

\section{Introduction}
Inflation solves the standard cosmological problems such as the flatness, horizon and monopole problems \cite{Starobinsky:1980te,Guth:1980zm,Sato:1980yn,Linde:1981mu,Starobinsky:1982ee,Albrecht:1982wi,Linde:1983gd}, and
quantum fluctuations of the inflaton during inflation seed the large-scale structure formation and leave imprints on the cosmic microwave background 
which can be tested by observations \cite{Smoot:1992td,Spergel:2003cb,Komatsu:2010fb,Aghanim:2018eyx,Akrami:2018odb,Sato:2015dga}. 
The Planck 2018 constraints on the amplitude $\mathcal{A}_\zeta$ and the spectral index $n_{\mathrm{s}}$ of the power spectrum of the primordial
curvature perturbations $\zeta$ are 
$\ln[10^{10}\mathcal{A}_\zeta] = 3.044\pm{0.014}$ and $n_\mathrm{s} = 0.9649 \pm 0.0042$ \cite{Akrami:2018odb}, 
and the non-detection of primordial gravitational waves limits the tensor-to-scalar ratio to be
$r_{0.05}<0.06$ (95\% C.L.) \cite{Akrami:2018odb,Ade:2018gkx}.

Since two-point correlation functions describe freely propagating particles in cosmological background,
so the information drawn from the power spectra, 
the Fourier transforms of the two-point correlation functions is limited \cite{Chen:2010xka,Bartolo:2004if,Seery:2005wm}.  
To obtain information about interactions and to break the degeneracy of inflationary models, 
higher order correlation functions, i.e. 
the non-Gaussianities of the primordial curvature perturbations need to be considered \cite{Salopek:1990jq,Gangui:1993tt,Komatsu:2001rj,Maldacena:2002vr,Creminelli:2004yq,
Creminelli:2005hu,Creminelli:2006rz,Chen:2006nt,Sreenath:2014nca,DeFelice:2011uc,
Kobayashi:2011pc,DeFelice:2011zh}. While we need only the quadratic action to calculate the two-point correlation functions, we need at least cubic-order action to include interaction terms which are responsible for three-point and higher-order correlation functions.
For canonical single field inflation under slow-roll conditions,
the non-Gaussianity is negligible \cite{Maldacena:2002vr}. 
If a significant non-Gaussianity is observed, then the class of
canonical single field slow-roll inflationary models can be ruled out.
The Planck 2018 constraints on the non-Gaussianity parameter are
$f^\text{local}_{\text{NL}}=-0.9\pm 5.1$, $f^\text{equil}_\text{NL}=-26\pm 47$
and $f^\text{ortho}_\text{NL}=-38\pm 24$ at the pivotal scale $k_*=0.05$ Mpc$^{-1}$ \cite{Akrami:2019izv}.

To get large non-Gaussianities, we need to consider inflationary models with non-canonical fields, 
or models violating the slow-roll conditions. 
An interesting class of models with the enhancement of the power spectrum of primordial curvature perturbations at small scales
attracted a lot of attentions recently \cite{Gong:2017qlj,Garcia-Bellido:2017mdw,Germani:2017bcs,Motohashi:2017kbs,
Ezquiaga:2017fvi,Bezrukov:2017dyv,Espinosa:2017sgp,Ballesteros:2017fsr,Ballesteros:2018wlw,
Sasaki:2018dmp,Kamenshchik:2018sig,Gao:2018pvq,Dalianis:2018frf,
Dalianis:2019vit,Passaglia:2018ixg,Ballesteros:2019hus,Passaglia:2019ueo,Fu:2019ttf,
Fu:2019vqc,Xu:2019bdp,Braglia:2020eai,Gundhi:2020zvb,Lin:2020goi,Yi:2020kmq,
Yi:2020cut,Gao:2020tsa,Fumagalli:2020adf,Fumagalli:2020nvq,Gundhi:2020kzm,Ballesteros:2020qam,Ragavendra:2020sop}. 
With the enhancement of the power spectrum at small scales, significant abundances of primordial black holes (PBHs) which may be the candidate of dark matter (DM) can be produced when the perturbations reenter the horizon during radiation domination \cite{Carr:1974nx,Hawking:1971ei,Ivanov:1994pa,Frampton:2010sw,Belotsky:2014kca,
Khlopov:2004sc,Clesse:2015wea,Carr:2016drx,Pi:2017gih,Inomata:2017okj,
Garcia-Bellido:2017fdg,Kovetz:2017rvv,Carr:2020xqk}, accompanied by the generation of scalar induced secondary gravitational waves (SIGWs)
which consist of the stochastic gravitational-wave background  \cite{Matarrese:1997ay,Mollerach:2003nq,Ananda:2006af,Baumann:2007zm,
Garcia-Bellido:2017aan,Saito:2008jc,Saito:2009jt,Bugaev:2009zh,Bugaev:2010bb,
Alabidi:2012ex,Orlofsky:2016vbd,Nakama:2016gzw,Inomata:2016rbd,Cheng:2018yyr,
Cai:2018dig,Bartolo:2018rku,Bartolo:2018evs,Kohri:2018awv,Espinosa:2018eve,
Kuroyanagi:2018csn,Cai:2019amo,Cai:2019elf,Cai:2019bmk,Cai:2020fnq,
Domenech:2019quo,Drees:2019xpp,Inomata:2019ivs,Inomata:2019zqy,
DeLuca:2019llr,Domenech:2020kqm,Braglia:2020taf}. The observations of both PBH DM and SIGWs can be used to test the models.
One particular class of models is  G-inflation model \cite{Kobayashi:2010cm,Kobayashi:2011nu} with a peak function in the non-canonical kinetic term \cite{Lin:2020goi,Yi:2020kmq,Yi:2020cut,Gao:2020tsa}.
The model can generate both sharp and broad peaks in the power spectrum without much fine-tuning of the model parameters.
To enhance the power spectrum at small scales and produce abundant
PBH DM and observational SIGWs, 
the violation of slow-roll conditions is inevitable. 
Therefore, the primordial non-Gaussianity in the models may possess different features which can be taken as a useful tool to discriminate them,
and large non-Gaussianities may arise at small scales. 
Then the impact of large non-Gaussianities on the abundance of PBHs and the power spectrum of SIGWs needs to be accounted for.

Based on this motivation, we study the primordial non-Gussianities from the G-inflation model. 
The bispectra, 
the Fourier transforms of three-point correlation functions ${\langle \zeta \zeta \zeta\rangle}$ of the curvature perturbations, 
represent the lowest order statistics able to distinguish non-Gaussian from Gaussian perturbations \cite{Bartolo:2004if}. We numerically compute  the bispectra and the relevant non-Gaussianity parameter $f_{\mathrm{NL}}(k_1,k_2,k_3)$ in the equilateral limit $k_1=k_2=k_3$, 
and the squeezed limit $k_3 \ll k_1=k_2$. 
In the squeezed limit, the bispectrum is related to the power spectrum by the consistency relation \cite{Maldacena:2002vr,Creminelli:2004yq} for single field inflation, 
no matter whether the slow-roll conditions are violated or not. 
It can be taken as a check of our numerical computation. 

This paper is organized as follows. In section II, 
we briefly review the calculation of non-Gaussianity first, 
then we present our results. 
We also discuss the effects of non-Gaussianities on the SIGWs and PBHs. 
The conclusions are drawn in section III.

\section{Non-Gaussianity from G-inflation}

The non-Gaussianity is negligible for a canonical single field inflation under slow-roll conditions \cite{Maldacena:2002vr,Bartolo:2004if,Chen:2010xka}. 
To get a large non-Gaussianity, a natural and feasible way is to break the slow roll conditions.
In this section, 
we study the non-Gaussianity from a class of generalized G-inflation models with a peak function in the non-canonical kinetic term.

\subsection{Non-Gaussianity and G-inflation}
The action of G-inflation model with a non-canonical kinetic term is \cite{Lin:2020goi,Yi:2020kmq,Yi:2020cut}
\begin{equation}\label{act:kgmodel}
S=\int d^4 x \sqrt{-g}\left[\frac{R}{2}+[1+G(\phi)]X-V(\phi)\right],
\end{equation}
where $X=-\nabla_{\mu}\phi\nabla^{\mu}\phi/2$, $G(\phi)$ is a general function with a peak, 
$V(\phi)$ is the inflationary potential and we choose $8\pi G=1$.
The non-canonical term $G(\phi)$ appears in Brans-Dicke theory of gravity,
k-inflation and G-inflation.

From the action \eqref{act:kgmodel}, we get the background equations
\begin{gather}
\label{Eq:eom1}
3H^2=\frac{1}{2}\dot{\phi}^2+V(\phi)+\frac{1}{2}\dot{\phi}^2G(\phi),\\
\label{Eq:eom2}
\dot{H}=-\frac{1}{2}[1+G(\phi)]\dot{\phi}^2,\\
\label{Eq:eom3}
\ddot{\phi}+3H\dot{\phi}+\frac{V_{\phi}+\dot{\phi}^2G_{\phi}/2}{1+G(\phi)}=0,
\end{gather}
where the Hubble parameter $H=\dot{a}/a$, $G_\phi=dG(\phi)/d\phi$ and $V_\phi=dV/d\phi$.

Perturbing the action \eqref{act:kgmodel} to the second order,
we get the quadratic action that determines the evolution of comoving curvature 
perturbation $\zeta$
\begin{equation}\label{S2}
S_2=\frac{1}{2} \int d \tau d^{3} x z^2\left[\left(\zeta^{\prime}\right)^{2}-(\partial \zeta)^{2}\right],
\end{equation}
where $z^2=2a^2\epsilon$, the prime denotes the derivative with respect to the conformal time ${\tau}$, $d\tau=a(t) dt$ and the slow-roll parameters $\epsilon$ and $\eta$ are
\begin{equation}\label{SRPA}
\epsilon=-\frac{\dot{H}}{H^{2}},\ \eta=\frac{\dot \epsilon}{H \epsilon}.
\end{equation}
Varying the quadratic action with respect to ${\zeta}$, we obtain the Mukhanov-Sasaki equation \cite{Mukhanov:1985rz,Sasaki:1986hm}
\begin{equation}
\label{E1}
\zeta_{k}^{\prime \prime}+2\frac{z'}{z}\zeta_{k}^{\prime}+k^2\zeta_k=0.
\end{equation}
Solving the above Mukhanov-Sasaki equation \eqref{E1}, 
we can calculate the two-point correlation function and the power spectrum as follows
\begin{equation}
\begin{split}
\langle \hat{\zeta}_{\bm{k}}\hat{\zeta}_{\bm{k}'}\rangle&=
(2\pi)^3\delta^3\left(\bm{k}+\bm{k}'\right)|\zeta_k|^2\\
&=
(2\pi)^3\delta^3\left(\bm{k}+\bm{k}'\right)P_{\zeta}\left(k\right),
\end{split}
\end{equation}
where the mode function $\zeta_k$ is the solution to Eq. \eqref{E1}, $\hat{\zeta}_{\bm{k}}$ is the quantum field of the curvature perturbation, the vacuum is chosen to be Bunch-Davies vacuum and the power spectrum is evaluated at late times when relevant modes are frozen.
The dimensionless scalar power spectrum and the scalar power index are defined as
\begin{equation}
\Delta^{2}_{\zeta}=\frac{k^3}{2\pi^2}P_\zeta\left(k\right),
\end{equation}
\begin{equation}
    n_{\mathrm{s}}(k)-1=\frac{d\mathrm{ln}\Delta ^2_{\zeta}}{d\mathrm{ln}k}.
\end{equation}

The bispectrum $B_{\zeta}$ is related with the three-point function as \cite{Byrnes:2010ft,Ade:2015ava} 
\begin{equation}\label{Bi}
\left\langle\hat{\zeta}_{\bm{k}_{1}}\hat{\zeta}_{\bm{k}_{2}}\hat{\zeta}_{\bm{k}_{3}}\right\rangle=(2 \pi)^{3} \delta^{3}\left(\bm{k}_{1}+\bm{k}_{2}+\bm{k}_{3}\right) B_{\zeta}\left(k_{1}, k_{2}, k_{3}\right),
\end{equation}
where the bispectrum $B_\zeta(k_1,k_2,k_3)$ is \cite{Hazra:2012yn,Ragavendra:2020old,Arroja:2011yj}
\begin{equation}
\label{bkeq1}
B_{\zeta}(k_1,k_2,k_3)=\Sigma^{10}_{i=1}B^{i}_{\zeta}(k_1,k_2,k_3),
\end{equation}
\begin{equation}\label{B1}
\begin{split}
B^{1}_{\zeta}(k_1,k_2,k_3)=&
-4\operatorname{Im}\left[\zeta_{{k}_{1}}(\tau_*)\zeta_{{k}_{2}}(\tau_*) \zeta_{{k}_{3}}(\tau_*) \int_{\tau_i}^{\tau_{*}}d\tau a^2\epsilon^{2}\left\{\zeta^{*}_{{k}_{1}}(\tau)\zeta^{'*}_{{k}_{2}}(\tau)\zeta^{'*}_{{k}_{3}}(\tau)
+\mathrm{perm}\right\}\right],
\end{split}
\end{equation}

\begin{equation}
\begin{split}
B^{2}_{\zeta}(k_1,k_2,k_3)=&
2\operatorname{Im}\left[
\vphantom{\left(\frac{\bm{k}_{2} \cdot \bm{k}_{3}}{k_{3}}\right)^{2}}
\zeta_{{k}_{1}}(\tau_*)\zeta_{{k}_{2}}(\tau_*) \zeta_{{k}_{3}}(\tau_*) \right.\\& \left. \int_{\tau_i}^{\tau_{*}}d\tau
a^2\epsilon^{2}\left\{(k^2_1-k^2_2-k^2_3)\zeta^{*}_{{k}_{1}}(\tau)\zeta^{*}_{{k}_{2}}(\tau)\zeta^{*}_{{k}_{3}}(\tau)
+\mathrm{perm}\right\}\right],
\end{split}
\end{equation}

\begin{equation}
\begin{split}
B^{3}_{\zeta}(k_1,k_2,k_3)=&
2\operatorname{Im}\left[
\vphantom{\left(\frac{\bm{k}_{2} \cdot \bm{k}_{3}}{k_{3}}\right)^{2}}
\zeta_{{k}_{1}}(\tau_*)\zeta_{{k}_{2}}(\tau_*) \zeta_{{k}_{3}}(\tau_*) \right. \\& \left. \int_{\tau_i}^{\tau_{*}}d\tau
a^2\epsilon^{2}\left\{\left(\frac{k^2_2-k^2_1-k^2_3}{k^2_1}+\frac{k^2_1-k^2_2-k^2_3}{k^2_2}\right)
\zeta^{'*}_{{k}_{1}}(\tau)\zeta^{'*}_{{k}_{2}}(\tau)\zeta^{*}_{{k}_{3}}(\tau)+\mathrm{perm}
\right\}\right],
\end{split}
\end{equation}

\begin{equation}
B^{4}_{\zeta}(k_1,k_2,k_3)=
-2\operatorname{Im}\left[\zeta_{{k}_{1}}(\tau_*)\zeta_{{k}_{2}}(\tau_*) \zeta_{{k}_{3}}(\tau_*) \int_{\tau_i}^{\tau_{*}}d\tau a^2\epsilon \eta'\left\{\zeta^{*}_{{k}_{1}}(\tau)\zeta^{*}_{{k}_{2}}(\tau)\zeta^{'*}_{{k}_{3}}(\tau)+\mathrm{perm}\right\}\right],
\end{equation}

\begin{equation}
\begin{split}
B^{5}_{\zeta}(k_1,k_2,k_3)=&
-\frac{1}{2}\operatorname{Im}\left[
\vphantom{\left(\frac{\bm{k}_{2} \cdot \bm{k}_{3}}{k_{3}}\right)^{2}}
\zeta_{{k}_{1}}(\tau_*)\zeta_{{k}_{2}}(\tau_*)\zeta_{{k}_{3}}(\tau_*) \right.\\& \left.
\int_{\tau_i}^{\tau_{*}}d\tau a^2\epsilon^{3}\left\{\left(\frac{k^2_2-k^2_1-k^2_3}{k^2_1}+\frac{k^2_1-k^2_2-k^2_3}{k^2_2}\right)
\zeta^{'*}_{{k}_{1}}(\tau)\zeta^{'*}_{{k}_{2}}(\tau)\zeta^{*}_{{k}_{3}}(\tau)
+\mathrm{perm}\right\}\right],
\end{split}
\end{equation}

\begin{equation}
\begin{split}
B^{6}_{\zeta}(k_1,k_2,k_3)=&
-\frac{1}{2}\operatorname{Im}\left[
\vphantom{\left(\frac{\bm{k}_{2} \cdot \bm{k}_{3}}{k_{3}}\right)^{2}}
\zeta_{{k}_{1}}(\tau_*)\zeta_{{k}_{2}}(\tau_*) \zeta_{{k}_{3}}(\tau_*)\right. \\& \left.
\int_{\tau_i}^{\tau_{*}}d\tau a^2\epsilon^{3}\left\{\frac{k^2_3 \left(k^2_3-k^2_1-k^2_2\right)}{k^2_1k^2_2}
\zeta^{'*}_{{k}_{1}}(\tau)\zeta^{'*}_{{k}_{2}}(\tau)\zeta^{*}_{{k}_{3}}(\tau)
+\mathrm{perm}\right\}\right],
\end{split}
\end{equation}

\begin{equation}
B^{7}_{\zeta}(k_1,k_2,k_3)=2\operatorname{Im}\left[\zeta_{k_{1}}(\tau_*) \zeta_{k_{2}}(\tau_*) \zeta_{k_{3}}(\tau_*)
\left(a^{2} \epsilon \eta \zeta_{k_{1}}^{*}(\tau) \zeta_{k_{2}}^{*}(\tau) \zeta_{k_{3}}^{\prime *}(\tau)+\mathrm{perm}\right)\right]\Big|_{\tau_i}^{\tau_*},
\end{equation}

\begin{equation}
\begin{split}
B^{8}_\zeta(k_1,k_2,k_3)=&2\operatorname{Im}\left[\zeta_{k_1}(\tau_*) \zeta_{k_2}(\tau_*) \zeta_{k_3}(\tau_*)\times\left(\frac{a}{H} \zeta_{k_{1}}^{*}(\tau) \zeta_{k_2}^{*}(\tau) \zeta_{k_{3}}^{*}(\tau)\right) \right.\\& \left.
\times \left\{54(a H)^{2}+2(1-\epsilon)(\bm{k}_1 \cdot \bm{k}_2+\mathrm{perm}) 
+ \right. \right. \\& \left. \left.  \frac{1}{2(a H)^{2}}\left[\left(\bm{k}_1 \cdot \bm{k}_2\right) k_{3}^{2}+\mathrm{perm}\right]\right\}\right] \Big|_{\tau_i}^{\tau_*},
\end{split}
\end{equation}

\begin{equation}
\begin{split}
B^{9}_{\zeta}(k_1,k_2,k_3)=&-\operatorname{Im}\left[\vphantom{\left(\frac{\bm{k}_{2} \cdot \bm{k}_{3}}{k_{3}}\right)^{2}} \zeta_{k_{1}}(\tau_*) \zeta_{k_{2}}(\tau_*) \zeta_{k_{3}}(\tau_*) 
 \left\{\frac{\epsilon}{ H^{2}} \zeta_{k_{1}}^{*}(\tau) \zeta_{k_{2}}^{*}(\tau) \zeta_{k_{3}}^{\prime *}(\tau) \right. \right. \\& \left.\left.
\left[k_{1}^{2}+k_{2}^{2}
-\left(\frac{\bm{k}_{1} \cdot \bm{k}_{3}}{k_{3}}\right)^{2}
-\left(\frac{\bm{k}_{2} \cdot \bm{k}_{3}}{k_{3}}\right)^{2}\right] +  \mathrm{perm}\right\}\right]\Big|^{\tau_*}_{\tau_i},
\end{split}
\end{equation}

\begin{equation}\label{B10}
\begin{split}
B^{10}_{\zeta}(k_1,k_2,k_3)=&2\operatorname{Im}\left[\vphantom{\left(\frac{\bm{k}_{2} \cdot \bm{k}_{3}}{k_{3}}\right)^{2}} \zeta_{k_{1}}(\tau_*) \zeta_{k_{2}}(\tau_*) \zeta_{k_{3}}(\tau_*) \right.\\& \left.
\times \left\{\frac{a\epsilon}{H} \zeta_{k_{1}}^{*}(\tau) \zeta_{k_{2}}^{\prime *}(\tau) \zeta_{k_{3}}^{\prime *}(\tau)
\left[2-\epsilon+\epsilon\left(\frac{\bm{k}_2\cdot \bm{k}_3}{k_2 k_3}\right)^2\right] +\mathrm{perm}\right\}\right]\Big|^{\tau_*}_{\tau_i},
\end{split}
\end{equation}
$\tau_*$ is a sufficient late time when all relevant modes have been frozen and $\tau_i$ is an early time when the plane wave initial conditions are imposed on the modes.
Because the modes oscillate rapidly well inside the horizon, its averaged contribution is negligible. 
However, in numerical computation, the choice of $\tau_i$ brings considerable uncertainty in the integral.
Following Refs.  \cite{Chen:2006xjb,Chen:2008wn}, 
we introduce a cutoff $\text{e}^{\lambda k_m (\tau-\tau_0)}$ in the integral to damp out the oscillatory contributions from very early times, where $k_m$ is the largest of the three modes, $\tau_0$ is a time several $e$-folds before $\tau_c$ when the $k_m$ mode crosses the horizon and $\lambda$ serves to determine to what extent the integral will be suppressed. Besides, $\lambda$ should satisfy $|\lambda k_m \tau_i| \gg 1$ and $|\lambda k_m \tau_c| \ll 1$ in order not to suppress the contribution from interactions. The existence of this cutoff also makes the result insensitive to the lower limit of the integration $\tau_i$ as long as the initial conditions are satisfied.

With the result of the bispectrum \eqref{bkeq1}, the non-Gaussianity parameter $f_\text{NL}$ is defined as \cite{Creminelli:2006rz,Byrnes:2010ft}
\begin{equation}\label{Fnl}
f_{\text{NL}} (k_1,k_2,k_3)=\frac{5}{6}\frac{B_{\zeta}(k_1,k_2,k_3)}{P_{\zeta}(k_1)
P_{\zeta}(k_2)+P_{\zeta}(k_2)P_{\zeta}(k_3)+P_{\zeta}(k_3)P_{\zeta}(k_1)}.
\end{equation}

\subsection{G-inflation with power-law potential}
\begin{figure}[htp]
  \centering
  \includegraphics[width=0.45\textwidth]{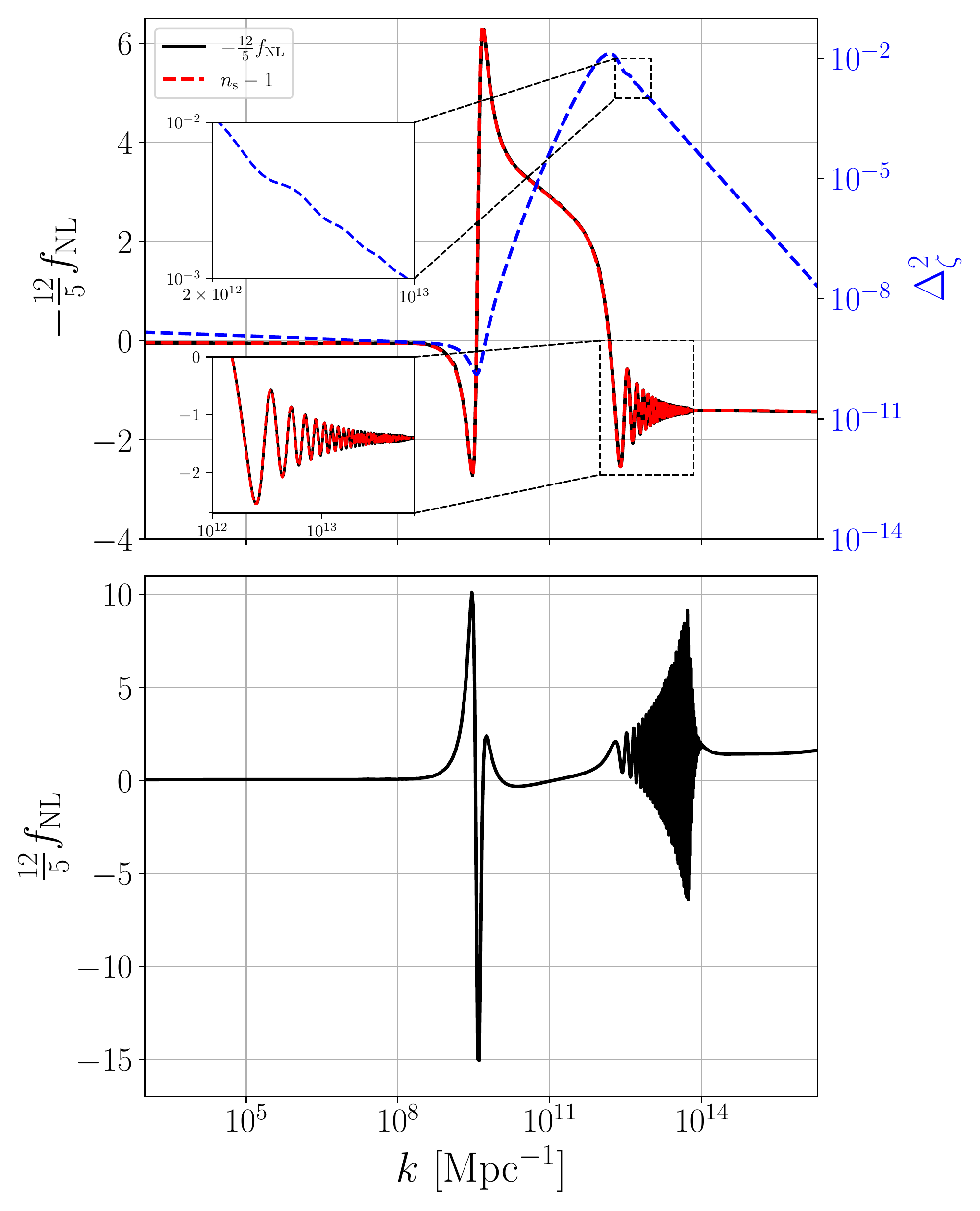}
  \caption{The primordial scalar power spectrum $\Delta_\zeta^2$ and the non-Gaussianity parameter $f_{\mathrm{NL}}$ for the G1 model \eqref{kg:peak} with the power-law potential $\lambda \phi^{2/5}$. The upper panel shows the power spectrum with the blue dashed line, $-\frac{12}{5}f_{\mathrm{NL}}$ in the squzeed limit with the solid black line for the modes $k_1=k_2=10^6 k_3=k$ and the scalar spectral tilt $n_\mathrm{s}-1$ with the red dashed line. The insets show the oscillations in $\Delta_\zeta^2$ and $f_{\mathrm{NL}}$. The lower panel shows $\frac{12}{5}f_{\mathrm{NL}}$ in the equilateral limit for the modes $k_1=k_2=k_3=k$.}\label{kgfnlps}
\end{figure}

In the G-inflation model \cite{Lin:2020goi}, the non-canonical kinetic term with a peak enhances the power spectrum on small scales so that abundant PBH DM and observable SIGWs are produced. 
The non-canonical kinetic function with a peak is \cite{Lin:2020goi}
\begin{equation}\label{kg:peak}
  G (\phi)=\frac{d}{1+|\frac{\phi-\phi_p}{c}|},
\end{equation}
and the chosen potential is $V(\phi)=\lambda \phi^{2/5}$, we call this model as G1. The peak function is inspired by the coupling $1/\phi$ in Brans-Dikce theory of gravity.
With the setting of parameters listed in Table. \ref{kg:tab} \cite{Lin:2020goi}, 
we numerically solve the background Eqs. \eqref{Eq:eom1}-\eqref{Eq:eom3} and the
Mukhanov-Sasaki Eq. \eqref{E1} to get the power spectrum.
We plot the power spectrum $\Delta^2_{\zeta}$ in the upper panel of Fig. \ref{kgfnlps}.
\begin{table}[htp]
    \centering
	\renewcommand\tabcolsep{4.0pt}
	\begin{tabular}{llllllllllll}
		\hline
		\hline
	Model &	$d$ & $c$ & $\phi_p$& $\phi_*$ & $\lambda$\\
		\hline
    G1  &  $5.26\times 10^{8}$&  $1.472\times 10^{-10}$ & $2.97$ & $ 5.21$ & $7.2\times 10^{-10}$\\
		\hline
		\hline
	\end{tabular}
	\caption{The parameters for the peak function \eqref{kg:peak} and the power-law potential $\lambda\phi^{2/5}$. $\phi_*$ is the value of the inflaton when the pivot scale $k_*=0.05\  \text{Mpc}^{-1}$ leaves the horizon.}
\label{kg:tab}
\end{table}

 With Eqs. \eqref{B1}-\eqref{Fnl}, we numerically compute the non-Gaussianity parameter $f_{\mathrm{NL}}$ in the squeezed limit and the equilateral limit. The results are shown in Fig. \ref{kgfnlps}. 

In the upper panel, we plot $-12f_{\mathrm{NL}}(k)/5$ along with the scalar spectral index $n_{\mathrm s}(k)-1$ to check the consistency relation
\begin{equation}\label{consistency}
    \lim_{k_3\rightarrow 0} f_{\mathrm{NL}}(k_1,k_2,k_3)=\frac{5}{12}(1-n_{\mathrm s}),\quad \text{for}~ k_1=k_2.
\end{equation}
Our results confirm that the consistency relation  \eqref{consistency} holds for models with peaks in the non-canonical kinetic term.
The consistency relation which relates a three-point correlation function to a two-point correlation function, was derived originally in the canonical single-field inflation with slow roll \cite{Maldacena:2002vr}. 
Then it was proved to be true for any inflationary model as long as the inflaton is the only dynamical field in addition to the gravitational field during inflation \cite{Creminelli:2004yq}.  
From our results, we see that the non-Gaussianity parameter $f_{\text{NL}}$ can reach as large as order 1, at certain scales,  $\left|12f_{\mathrm{NL}}\right|/5\simeq 6$ in the squeezed limit and $\left|12f_{\mathrm{NL}}\right|/5\simeq 15$ in the equilateral limit. 
Both maxima for $f_{\text{NL}}$ happen at the scales where
the power spectrum has a small dip before it reaches the peak.
For the model G1 considered in this subsection, there is a sharp spike for $f_{\text{NL}}$ at the scales between $k\sim 10^9~\mathrm{Mpc}^{-1}$ and $k\sim 10^{11}~\mathrm{Mpc}^{-1}$. 
The oscillation nature of $f_{\text{NL}}$ at scales around $k\sim 10^{13}~\mathrm{Mpc}^{-1}$ in both squeezed and equilateral limits originates from the small waggles in the power spectrum, 
as can be seen from the inset of Fig. \ref{kgfnlps}. 
The oscillation in the power spectrum results in the oscillation in $f_{\mathrm{NL}}$ in the equilateral limit was also studied in Ref. \cite{Hazra:2012yn}.

If an inflationary model produces a power spectrum with a large slope, i.e., the power spectral $|n_s-1|\gg 1$, then the model has a large non-Gaussianity, at least in the squeezed limit, as told by the consistency relation \eqref{consistency} and shown in Fig. \ref{kgfnlps}. 

In terms of the nonlinear coupling constant $f_{\text{NL}}$ defined as \cite{Verde:1999ij,Komatsu:2001rj}
\begin{equation}\label{local}
\zeta(\bm{x})=\zeta^G(\bm{x})+\frac{3}{5}f_\text{NL}(\zeta^G(\bm{x})^2-\langle \zeta^G(\bm{x})^2 \rangle),
\end{equation}
where $\zeta^G$ is the linear Gaussian part of the curvature perturbation, 
it was shown that the non-Gaussian contribution to SIGWs can exceed the Gaussian part if  
$\left(3/5\right)^2f^2_{\text{NL}} \Delta^2_\zeta\gtrsim 1$ \cite{Cai:2018dig}.
Taking $f_\mathrm{NL}$ as the estimator to parameterize the magnitude of non-Gaussianity \cite{Chen:2006nt}, we expect the contribution from the non-Gaussian scalar perturbations to SIGWs is negligible because $f_{\text{NL}}$
is small when the peak is reached as shown in Fig. \ref{kgfnlps}.

The effect of non-Gaussianity of the curvature perturbation $\zeta$ on the PBH
abundance is characterized by the factor $\mathcal{J}$ \cite{Seery:2006wk,Hidalgo:2007vk,Saito:2008em}
\begin{equation}
\label{jeq1}
\mathcal{J}=\frac{1}{6 \sigma^3_R} \int \frac{\mathrm{d}^{3} k_{1}}{(2\pi)^3} \int \frac{\mathrm{d}^{3} k_{2}}{(2\pi)^3} \int \frac{\mathrm{d}^{3} k_{3}}{(2\pi)^3} \mathrm{W}\left(k_{1} R\right) \mathrm{W}\left(k_{2} R\right) \mathrm{W}\left(k_{3} R\right)\left\langle\hat{\zeta}_{k_{1}} \hat{\zeta}_{k_{2}} \hat{\zeta}_{k_{3}}\right\rangle,
\end{equation}
where the variance
\begin{equation}
\label{sigreq1}
\sigma_R^2=\int \frac{dk}{k} \mathrm{W}(kR)^2 \Delta_\zeta^2,
\end{equation}
and $\mathrm{W}(kR)$ is a window function. If $\mathcal{J}\ll 1$,
then the contribution of non-Gaussianity of the curvature perturbation to PBH abundance is negligible.
Since the major contribution to $\mathcal{J}$ is around the peak of the power spectrum, 
for a good approximation we can use the value of $\mathcal{J}$ at the peak scale of the power spectrum of the curvature perturbation to evaluate the effect of non-Gaussianities on PBH abundance \cite{Saito:2008em}, 
\begin{equation}
\label{jeq2}
\mathcal{J}_\mathrm{peak}=\frac{3}{20\pi}f_\mathrm{NL}(k_\mathrm{peak},k_\mathrm{peak},k_\mathrm{peak})\sqrt{\Delta^2_\zeta(k_\mathrm{peak})}.
\end{equation}
From Fig. \ref{kgfnlps}, we see that $f_\mathrm{NL}(k_\mathrm{peak},k_\mathrm{peak},k_\mathrm{peak}) \sim \mathcal{O}(1)$ and $\Delta^2_\zeta (k_\mathrm{peak})\sim \mathcal{O}(0.01)$. 
Plugging these numbers into Eq. \eqref{jeq2}, we obtain
$\mathcal{J}_\mathrm{peak} \ll 1$, so we expect that the effect of non-Gaussianity on PBHs abundance is negligible.

\subsection{G-inflation with Higgs field}

For the G-inflation model G1 with the peak function \eqref{kg:peak} to be consistent with observational constraints, 
the potential is restricted to be $V(\phi)=\lambda \phi^{2/5}$. 
To lift the restriction on both the peak function and the inflationary potential, the non-canonical kinetic term in the model is generalized to \cite{Yi:2020kmq,Yi:2020cut}
\begin{equation}\label{higgs:peak}
    G(\phi)=G_p(\phi)+f(\phi),
\end{equation}
where the peak function $G_p$ is generalized to
\begin{equation}\label{GP}
G_p(\phi)=\frac{d}{1+|(\phi-\phi_p)/c|^q},
\end{equation}
the function $f(\phi)$ is used to dress the non-canonical scalar field. 
It was shown that the enhanced power spectrum can have both sharp and broad peaks with the peak function \eqref{GP}. 
In this mechanism, Higgs inflation, T-model and natural inflation are shown to satisfy the observational constraints at large scales and the amplitudes of the power spectra are enhanced  by seven orders of magnitude at small scales \cite{Yi:2020kmq,Yi:2020cut,Gao:2020tsa}.

\begin{figure}[htp]
  \centering
  \includegraphics[width=0.45\textwidth]{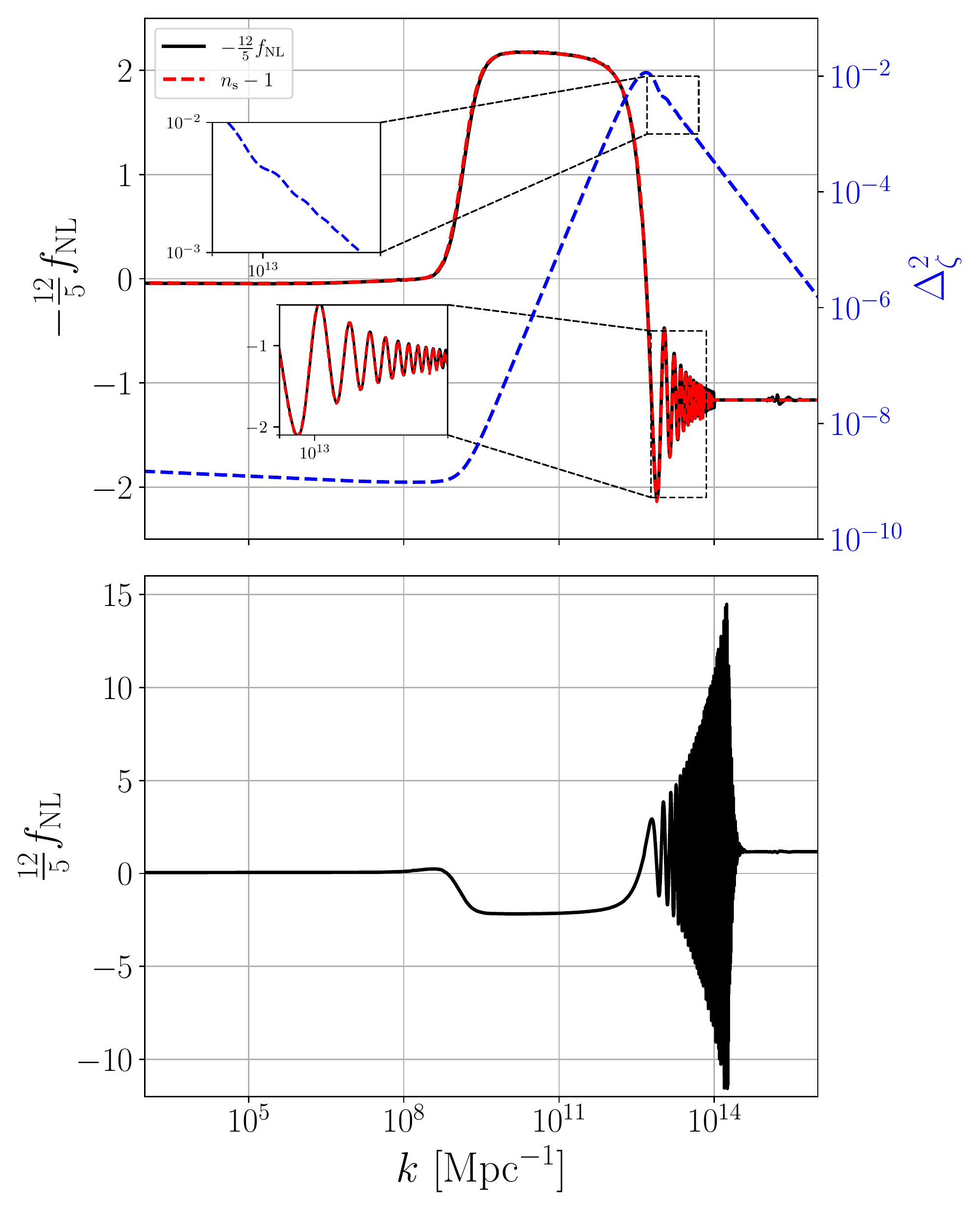}
  \caption{The primordial scalar power spectrum $\Delta_\zeta^2$ and the non-Gaussianity parameter $f_{\mathrm{NL}}$ for the Higgs model H11. The upper panel shows the power spectrum with the blue dashed line, $-\frac{12}{5}f_{\mathrm{NL}}$ in the squeezed limit with the solid black line for the modes $k_1=k_2=10^6 k_3=k$ and the scalar spectral tilt $n_\mathrm{s}-1$ with the red dashed line.
  The insets show the oscillations in $\Delta_\zeta^2$ and $f_{\mathrm{NL}}$. The lower panel shows $\frac{12}{5}f_{\mathrm{NL}}$ in the equilateral limit for the modes $k_1=k_2=k_3=k$.}\label{higgsfnlps}
\end{figure}

\begin{figure}[htp]
  \centering
  \includegraphics[width=0.450\textwidth]{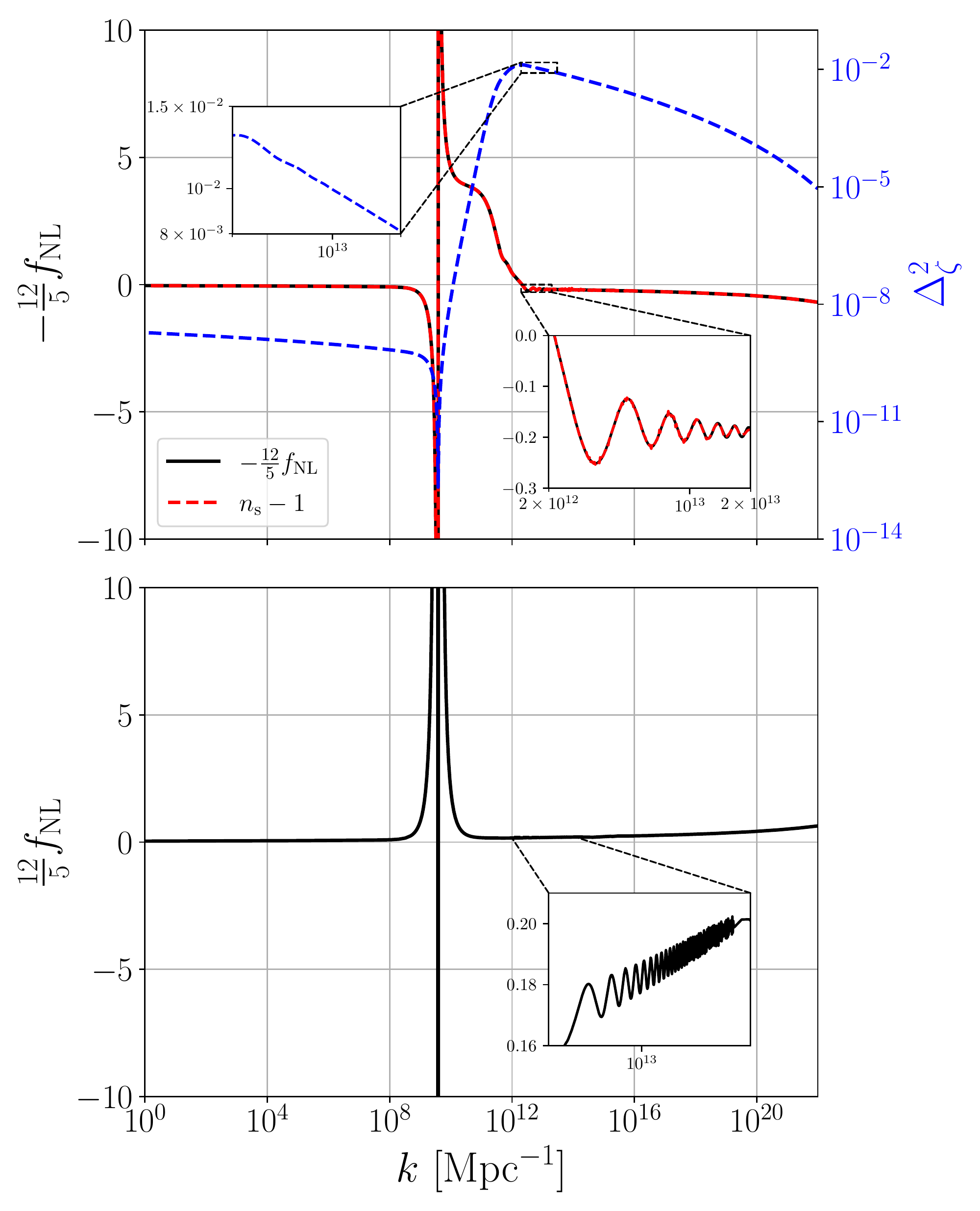}
  \caption{The primordial scalar power spectrum $\Delta_\zeta^2$ and the non-Gaussianity parameter $f_{\mathrm{NL}}$ for the Higgs model H12. The upper panel shows the power spectrum with the blue dashed line, $-\frac{12}{5}f_{\mathrm{NL}}$ in the squeezed limit with the solid black line for the modes $k_1=k_2=10^6 k_3=k$ and the scalar spectral tilt $n_\mathrm{s}-1$ with the red dashed line.
  The insets show the oscillations in $\Delta_\zeta^2$ and $f_{\mathrm{NL}}$. The lower panel shows $\frac{12}{5}f_{\mathrm{NL}}$ in the equilateral limit for the modes $k_1=k_2=k_3=k$.}\label{higgs12fnlps}
\end{figure}

For the Higgs field with the potential $V(\phi)=\lambda \phi^4 /4$, the function $f(\phi)$ is chosen to be \cite{Yi:2020kmq,Yi:2020cut}
\begin{equation}\label{higgs:fhiggs}
    f(\phi)=f_0 \phi^{22},
\end{equation} 
with $f_0$ a coupling constant. 
Following Ref. \cite{Yi:2020cut}, we use H11 to represent the model with $q=1$, and H12 to represent the model with $q=5/4$.
With the parameter settings as shown in Table. \ref{higgs:tab} \cite{Yi:2020kmq,Yi:2020cut}, we numerically compute the power spectrum $\Delta^2_{\zeta}$ and the non-Gaussianity parameters. We plot the results for the model H11 and the model H12 in Figs. \ref{higgsfnlps} and \ref{higgs12fnlps}, respectively.
\begin{table}[htp]
    \centering
	\renewcommand\tabcolsep{4.0pt}
	\begin{tabular}{lllllllllll}
		\hline
		\hline
		Model & $q$ & $d$ & $c$ & $\phi_p$ & $ \phi_*$ &  $f_0$ &  $\lambda$\\
		\hline
        H11 & $1$ & $1.05\times 10^{10}$&  $2.04\times 10^{-10}$ & $1.334$ & $1.40$ & $1$ &  $1.240\times 10^{-9}$\\
       \hline
        H12 & $5/4$ & $1.60\times 10^{11}$&  $2.06\times 10^{-10}$ & $1.300$ & $1.40$ & $1$ &  $1.244\times 10^{-9}$\\
		\hline
		\hline
	\end{tabular}
	\caption{The parameters for the peak function \eqref{GP} and the Higgs potential.}
\label{higgs:tab}
\end{table}


From the upper panels of Figs. \ref{higgsfnlps} and  \ref{higgs12fnlps}, 
we see that the peak of the power spectrum produced by H12 is broader than that produced by H11, 
this is because the peak function $G_p(\phi)$ of the model H12 is flatter than H11.

In the upper panels of Figs. \ref{higgsfnlps} and \ref{higgs12fnlps}, 
we also plot the non-Gaussianity parameter $-12f_\mathrm{NL}/5$ in the squeezed limit along with the scalar spectral index $n_\mathrm{s}(k)-1$ to check the consistency relation \eqref{consistency}. 
As shown in Figs. \ref{higgsfnlps} and \ref{higgs12fnlps}, 
the consistency relation still holds for models with either broad or sharp peaks in the non-canonical kinetic term, even when the slow roll conditions violate.

From Fig. \ref{higgsfnlps}, we see that in the model H11 with a sharp peak in the power spectrum, 
the non-Gaussianity parameter $f_\mathrm{NL}$ in the squeezed limit can be as large as $\left|12f_\mathrm{NL}/5\right|\simeq 2$ and $\left|12f_\mathrm{NL}/5\right|\simeq 15$ in the equilateral limit at certain scales. 
$f_\mathrm{NL}$ is almost a constant when the power spectrum climbs up to the peak. 
Due to the waggles in the power spectrum as shown in the insets of Fig. \ref{higgsfnlps},
$f_{\text{NL}}$ becomes large and it also shows 
the oscillation behaviour.

From Fig. \ref{higgs12fnlps}, we see that $f_{\mathrm{{NL}}}$ has a pretty large value in the model H12 with a broad peak in the power spectrum at scales around $k \sim 10^9\ \text{Mpc}^{-1}$ due to the plunge by several orders of magnitude in the power spectrum, and the amplitude is about $10^{-14}$ at the relevant scales.
However, $f_\mathrm{NL}$ is small at scales where the power spectrum reaches the peak. 
Again the waggles in the power spectrum as shown in the insets of Fig. \ref{higgs12fnlps} result in the oscillations in $f_{\mathrm{NL}}$.

Our numerical computations show that the three inflationary models predict different non-Gaussianities which could be taken as a tool to distinguish them. In general, there is much more information contained in the bispectrum than in the power spectrum.
By virtue of non-Gaussianity, we may differentiate inflationary models when degeneracy happens.

For all these models, $f_{\mathrm{NL}}$ is small at the scales where the power spectrum reaches the peaks, so we don't expect a large contribution from non-Gaussianity in the enhanced curvature perturbations to SIGWs and PBH abundance.
Even large non-Gaussianities arise in the model H12, the contribution to SIGWs and PBH abundance is still negligible because the amplitudes of the perturbations are very small at those scales. 

\section{Conclusion}

Non-Gaussianity is useful for understanding the dynamics of inflation and distinguishing different inflation models. 
We studied the non-Gaussianity of scalar perturbation produced in the G-inflation model with a peak function in the non-canonical kinetic term. 
The results confirm that the consistency relation holds in G-inflation models.
Even though the peak function enhances the scalar power spectrum by seven orders of magnitude at small scales, 
the non-Gaussianity is small at the peak scales. 
For the model G1,
$|12 f_{\text{NL}}/5|$ can be as large as 6 in the squeezed limit and 15 in the equilateral limit. 
Because of the dip before the peak scale in the power spectrum for the model G1, $f_{\text{NL}}$ reaches its maximum value at the dip scale.
The small waggles in the power spectrum not only result in the oscillation of $f_{\text{NL}}$, 
but also lead to large non-Gaussianity.
For the Higgs model H11, $|12f_{\text{NL}}/5|$ is almost an order 1 constant in both squeezed and equilateral limits at scales where the power spectrum climbs up the peak. 
At the peak scale, $f_{\text{NL}}$ is small. 
After the peak, the small wiggles in the power spectrum lead to the oscillation of $|12f_{\text{NL}}/5|$ with the amplitude of order 2 in the squeezed limit and the amplitude of 15 in the equilateral limit.
For the Higgs model H12,
the power spectrum plunges by several orders of magnitude before it reaches the broad peak and the non-Gaussianity becomes very large at the plunge scales. However, $f_{\text{NL}}$ keeps to be small at the peak scales.
Since $f_{\text{NL}}$ is small at the peak scales for all three models considered,
the contributions of the non-Gaussianity of curvature perturbations
to SIGWs and PBH abundance are negligible.

\begin{acknowledgments}
FZ would like to thank S. Passaglia for useful discussion.
This research was supported in part by the National Key Research and Development Program of China under Grant No. 2020YFC2201504;
the National Natural Science Foundation of China under Grant No. 11875136;
MOE Key Laboratory of TianQin Project, Sun Yat-sen University;
and the Major Program of the National Natural Science Foundation of China under Grant No. 11690021.
\end{acknowledgments}

\providecommand{\href}[2]{#2}\begingroup\raggedright\endgroup

\end{document}